\documentclass[aps,prx,twocolumn,superscriptaddress,floatfix]{revtex4-2}

\IfFileExists{supplement.pdf}{}{}

\usepackage{graphicx}
\usepackage{amsmath}
\usepackage{amssymb}
\usepackage{bm}

\usepackage{epsfig}
\usepackage{xcolor}
\usepackage{dutchcal}
\usepackage{hyperref}
\usepackage{soul}

\usepackage{placeins}

\hypersetup{
    colorlinks=true,
    citecolor=blue,
    urlcolor=blue,
    linkcolor=blue}

\begin{document}

\title{\textbf{Marrying Critical Oscillators with Traveling Waves \\ Shapes Nonlinear Sound Processing in the Cochlea}}

\author{Henri Ver Hulst}
\affiliation{Physics of Cells and Cancer, Institut Curie, PSL Research University, Sorbonne Université, CNRS UMR168, F-75248 Paris, France}

\author{Carles Blanch Mercader}
\affiliation{Physics of Cells and Cancer, Institut Curie, PSL Research University, Sorbonne Université, CNRS UMR168, F-75248 Paris, France}

\author{Frank Jülicher}
\affiliation{Max Planck Institute for the Physics of Complex Systems, 01187 Dresden, Germany}
\affiliation{Cluster of Excellence Physics of Life, Technische Universität, 01307 Dresden, Germany}

\author{Pascal Martin}
\email{pascal.martin@curie.fr}
\affiliation{Physics of Cells and Cancer, Institut Curie, PSL Research University, Sorbonne Université, CNRS UMR168, F-75248 Paris, France}

\date{\today}

\begin{abstract}
The cochlea's capacity to process a broad range of sound intensities has been linked to nonlinear amplification by critical oscillators. However, while the increasing sensitivity of a critical oscillator upon decreasing the stimulus magnitude comes with proportionally sharper frequency tuning and slower responsiveness---critical slowing down, the observed bandwidth of cochlear frequency tuning and the cochlear response time vary little with sound level. Because the cochlea operates as a distributed system rather than a single critical oscillator, it remains unclear whether criticality can serve as a fundamental principle for cochlear amplification. Here we tackle this challenge by integrating tonopically distributed critical oscillators in a traveling-wave model of the cochlea. Importantly, critical oscillators generically provide spatial buildup of energy gain from energy pumping into the waves and a key nonlinearity. In addition, our nonlinear model accounts for viscoelastic coupling between the oscillators. The model produces, with a single set of parameters, a family of cochlear tuning curves that quantitatively describe experimental data over a broad range of input levels. Overall, the interplay between generic nonlinear properties of local critical oscillators and distributed effects from traveling waves gives rise to a collective nonlinear response that preserves the power-law responsiveness afforded by criticality, but without paying the price of critical slowing down.
\end{abstract}

\maketitle

\section{Introduction}

Sound processing by the cochlea---the auditory receptor organ of the mammalian inner ear---is associated with an essential non-linearity~\cite{Robles2001,Goldstein1967}. From the faintest sounds detected to sounds so loud that they hurt, the sound-pressure level increases by six orders of magnitude, which elicit only two-to-three orders of magnitude of mechanical cochlear vibration. This compressive nonlinearity is widely thought to emerge from an active process that boosts sensitivity to weak sound stimuli, called the cochlear amplifier~\cite{Ashmore2010,Hudspeth2014}.

The nonlinear properties of the cochlear amplifier have been recognized as signatures of critical oscillators~\cite{Eguiluz2000,Duke2003,Kern2003,Magnasco2003,Hudspeth2010,Alonso2025}. A critical oscillator is an active dynamical system that operates at the onset of an oscillatory instability called a Hopf bifurcation~\cite{Strogatz1997,Crawford1991}. As expected for a critical oscillator (see generic properties in Section~B of the Supplemental Material~\cite{SuppMat}), the nonlinear relation between the sound-pressure level and the amplitude of basilar-membrane vibration in the cochlea can be captured by a power law with an exponent of about $+1/3$ over most of the range of sound-pressure levels. Because amplification by a critical oscillator is most effective for stimuli at the natural frequency of the oscillator, oscillatory inputs are also actively filtered, which sharpens frequency selectivity. Accordingly, cochlear sensitivity drops and tuning curves broaden when the active process is impaired~\cite{Sellick1982,Strimbu2022,Brown1983,Xia2022}. Finally, operating on the oscillatory side of the bifurcation may provide the substrate for the production of oto-acoustic emissions~\cite{Fruth2014,Vilfan2008}. In return, criticality has been proposed as a general principle to encapsulate in one concept disparate observations linked to auditory processing in the cochlea~\cite{Hudspeth2010,Hudspeth2024}.

While this principle offers an appealing simplification to describe nonlinear cochlear mechanics, the relevance of criticality remains debated~\cite{Lyon2017,Kondylidis2025,Shera2022}. One source of contention is a qualitative discrepancy in the relationship between sensitivity and the bandwidth of frequency tuning. A critical oscillator shows inverse level dependencies for its peak sensitivity, $\chi_{\max} \propto F^{-2/3}$, and the bandwidth of its tuning curve, $\Delta\omega \propto F^{2/3}$, with respect to the stimulus level, $F$. As a result, the product of these two quantities---called the gain-bandwidth product in the following---remains constant. The large sensitivity to stimuli of vanishing magnitudes is thus associated with tuning curves of vanishing bandwidth and, correspondingly, with long response times. This property is analogous to critical slowing down for thermodynamic phase transitions and is a defining feature of criticality. The cochlea does not work by this rule. As sound-pressure levels increase, the product of maximal sensitivity and bandwidth in a cochlear tuning curve decreases; while sensitivity to low sound levels is high, tuning remains relatively broad and increases little with level~\cite{Robles2001}. The cochlea thus appears to violate a fundamental property of sound detection by critical oscillators.

The response of the cochlea to a pure tone, however, cannot be reduced to that of a single oscillator. Instead, the cochlea functions as a distributed system, in which vibrations at one location along the basilar membrane influence other regions due to longitudinal hydrodynamic and mechanical interactions. As first demonstrated in the pioneering work of von Békésy~\cite{vonBekesy1960}, a pure-tone stimulus in the ear canal excites a wave of transverse basilar-membrane vibration that travels toward the apex of the cochlear tube. The wave progresses until it reaches a frequency-dependent characteristic place where the amplitude of the wave peaks before plummeting. The wave is elicited by air-pressure modulation in the ear canal, leading to changes in the local cross-sectional fluid pressure, which, in turn, drives the transverse motion of the basilar membrane at a given location within the cochlea. Local basilar-membrane vibrations are thus shaped by a complex interplay between local mechanical properties and distributed effects from the traveling wave. This interplay obscures our understanding of the cochlea and calls for a systematic study of traveling-wave models based on critical oscillators.

In this work, we marry tonotopically distributed critical oscillators and traveling waves in a box model of the cochlea. We examine how frequency-tuning curves are influenced by two-dimensional hydrodynamics, viscoelastic coupling between the oscillators, and energy pumping into the traveling wave. We find that our nonlinear model can quantitatively account for experimental tuning curves over a broad range of sond-pressure levels. It captures the generic power-law behavior associated with criticality, while producing broader tuning curves with a bandwidth that increases only weakly with level. Finally, we show that a minimal description of the cochlea based on one-dimensional hydrodynamics provides a simplified quantitative description of the cochlea, emphasizing the key roles of energy pumping and mechanical coupling to produce both sensitive and fast responses to faint sound stimuli.

\section{Results}

\subsection{Theoretical description of the cochlea}

The cochlear tube is split in two compartments by the organ of Corti---an epithelial strip that houses the mechanosensory hair cells and is supported by the basilar membrane. We developed a nonlinear two-dimensional (2D) box model of the cochlea in which the organ of Corti is represented by a string of oscillators (Fig.~\ref{fig:1}; see details in Section C of the Supplemental Material~\cite{SuppMat}). The natural angular frequencies of the oscillators are tuned according to an exponential tonotopic map, $\omega_{R}(x) = \omega_{0} \exp(-x/d)$, where $x$ is the position along the longitudinal axis of the cochlea, $\omega_{0}$ is the angular frequency of the oscillator at the base ($x = 0$) and $d$ is a characteristic length. The string of oscillators is positioned at $y = 0$ and divides the box in the $x$-$y$ plane into two identical compartments of height $H$ and length $L$, both filled with an incompressible and inviscid fluid (Fig.~\ref{fig:1}). The hydrodynamic interplay between the oscillators' motion and fluid flow is described by a standard approach~\cite{Lighthill1981,deBoer1996} and solved in the time domain (see methods in Section~D of the Supplemental Material~\cite{SuppMat}).

In this work, we followed a principle of criticality, by which each oscillator operates precisely at a Hopf bifurcation---all the oscillators are critical. Their equation of motion then obeys a generic equation of a complex variable, $z(x, t) = h(x, t) + i\, \text{Im}[z(x, t)]$, called the normal form, which was supplemented with mechanical coupling and external forcing:
\begin{equation} \label{eqMain:HopfNormalForm}
\partial_{t}z \simeq -(\varepsilon - i\omega_{R}(x))z - \beta|z|^{2}z + \kappa\partial_{x}^{2}z + \frac{e^{-i\phi}}{\alpha}p_{d}
\end{equation}

\begin{figure}[t]
\includegraphics[width=\columnwidth]{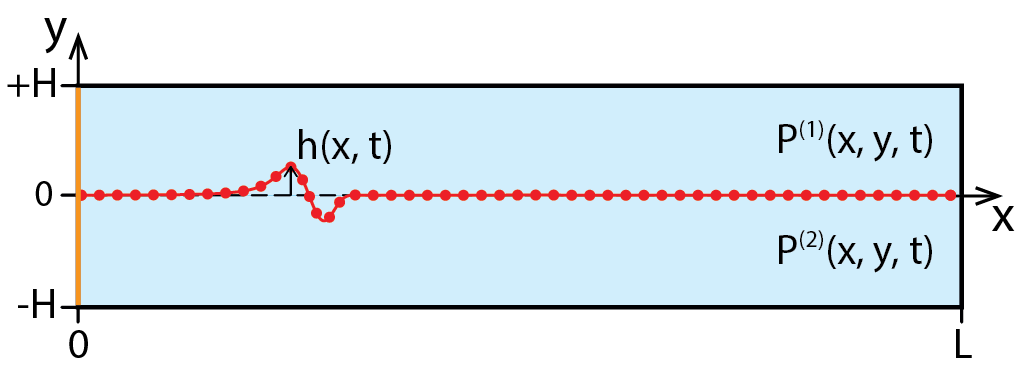} \label{figMain:boxModel}
\caption{Schematic of the box model. The cochlea is represented by a rectangular box of height $2H$ and length $L$. The box is filled by an incompressible and inviscid fluid (blue shading) and divided into two compartments of equal height by a string of critical oscillators (red disks). Their natural frequency, $\omega_{R}(x) = \omega_{0} \exp(-x/d)$ decreases exponentially from the base ($x = 0$) toward the apex ($x = L$) of the box. Under resting conditions, all the oscillators are positioned at $y = 0$ (black dashed line). Sound stimuli evoke a time and position dependent change of the fluid pressure in each compartment, $P^{(i)}(x, y, t)$ with $i = 1,2$. The pressure difference, $p_{d}(x, t) = P^{(1)}(x, y = 0, t) - P^{(2)}(x, y = 0, t)$, drives a transverse motion of the oscillators, $h(x, t)$. By this convention, a positive driving pressure evokes a downward movement. The stimulus is injected in the upper compartment at the open boundary positioned at the base ($x = 0$, orange line). We used $H = 1$ mm, $L = 19$ mm, $\omega_{0} = 1.89~\textrm{rad~s}^{-1}$ and $d = 2.9$~mm; other parameter values can be found in Table~S1 \cite{SuppMat}. The movements represented here have been magnified by a factor of about 15,000 with respect the maximal amplitude of 20 nm evoked by a pure tone of 6.6 kHz at a sound-pressure level of 40 dB in a chinchilla cochlea~\cite{Rhode2007}. Their spatial profile, however, is realistic.}
\label{fig:1}
\end{figure}

The value $\varepsilon = 0$ of the control parameter ensures criticality, the variable $h = \text{Re}(z)$ describes the basilar-membrane displacement, the complex coefficient $\beta = \beta_{r} + i\beta_{i}$ sets the strength of the oscillator's nonlinearity ($\beta_r>0$ to ensure stability), the coefficient $\alpha$ is real and sets the response magnitude at low frequencies (i.e. stiffness), $p_{d}(x, t)$ is the driving pressure at position $x$ and time $t$, and the parameter $\phi$ is a phase shift that arises from the change of variable to bring the system's dynamics into normal form~\cite{Julicher2009}. The third term on the right-hand side of Equation~\ref{eqMain:HopfNormalForm} accounts for longitudinal viscoelastic coupling between neighboring oscillators, with a coupling strength set by $\kappa = \kappa_{r} + i\kappa_{i}$. We use $\kappa_{r}(x) = \bar{\kappa}_{r}\omega_{R}(x)$, $\kappa_{i}(x) = \bar{\kappa}_{i}\omega_{R}(x)$, and $\alpha(x) = \bar{\alpha}\omega_{R}(x)$ to account for mechanical gradients along the tonotopic axis of the cochlea and, for simplicity, we assume that $\beta$ and $\phi$ do not depend on position.

We studied the response of the cochlear model to pure tones, $P_{\mathrm{EC}}(t) = P_{0}\sin(\omega t)$, in the ear canal at varying angular frequencies, $\omega$, and amplitude, $P_{0}$. The latter sets the sound-pressure level in decibels, $I_{dB} = 20\log_{10}(P_{0}/(\sqrt{2}P_{\mathrm{REF}}))$, with $P_{\mathrm{REF}} = 20$ $\mu$Pa. The input to the cochlear model is given by the pressure field at the base of the cochlea ($x = 0$), which is proportional to the product of the ear-canal pressure, $P_{\mathrm{EC}}(t)$, and the middle-ear gain, $G_{\mathrm{ME}}$ (elaborated in Section C3 of the Supplemental Material~\cite{SuppMat}).

We confronted the model to experimental measurements of basilar-membrane vibrations in the chinchilla cochlea at a fixed position of characteristic frequency $\textrm{CF} = 6.6$ kHz---the frequency at which sensitivity peaks at low levels (here 10 dB; cyan in Fig.~\ref{fig:2})~\cite{Rhode2007}. In the model, we thus analyzed the tuning curves of sensitivity, $\chi = |\tilde{h}/\tilde{P}_{\mathrm{EC}}|$, as a function of frequency, $f=\omega/(2\pi)$, at the fixed position, $x = \textrm{CP}$---the characteristic place in the model when driven by a stimulus at the characteristic frequency in experiments (elaborated in Sections C4 and D2.e of the Supplemental Material~\cite{SuppMat}). We varied the amplitude of sound pressure in the ear canal, $P_{0}$, and studied how the tuning curves changed shape. Here and in the following, Fourier transforms at the frequency of stimulation are denoted by tilde, e.g. $\tilde{h}(x,\omega) = \int dt\, h(x,t) e^{-i\omega t}$.

It is instructive to discuss the linear response of a critical oscillator to a pure tone at angular frequency $\omega$ in the absence of coupling ($\beta = 0$ and $\kappa = 0$ in Eq.~1; details in Section B2 of the Supplemental Material~\cite{SuppMat}). The complex modulus, $\mathcal{A} = \tilde{p}_{d}/\tilde{h}$, that characterizes the local linear response is given by
\begin{equation}
\mathcal{A}(x, \omega) = \alpha(x)\frac{\omega_{R}^{2}(x) - \omega^{2}}{\omega_{R}(x)\sin(\phi) + i\omega\cos(\phi)},
\end{equation}
whose imaginary part
\begin{equation} \label{eqMain:ComplexModulusImag}
\text{Im}[\mathcal{A}(x, \omega)] = -\alpha(x)\cos(\phi)\frac{\omega(\omega_{R}^{2}(x) - \omega^{2})}{\omega_{R}^{2}(x)\sin^{2}(\phi) + \omega^{2}\cos^{2}(\phi)},
\end{equation}
controls the time-averaged power transfer, $\mathcal{P} = \langle p_{d}(t) \partial_{t}h \rangle_{t} = 2\omega|\tilde{h}|^{2}\text{Im}[\mathcal{A}]$, from the environment to the oscillator. An important generic property of a critical oscillator is that it pumps energy into its environment, $\mathcal{P} < 0$, when driven at an angular frequency $\omega$ below its natural frequency $\omega_{R}$ if the parameter $\phi$ lies within the range $0 < \phi < \pi/2$. This behavior requires an active process and, thus, cannot happen in a passive system~\cite{Hudspeth2010,Martin2021}.

\subsection{A 1D model with no mechanical coupling nor energy pumping yields too-sharp tuning and level-independent gain-bandwidth product}

\begin{figure*}[t]
\includegraphics[]{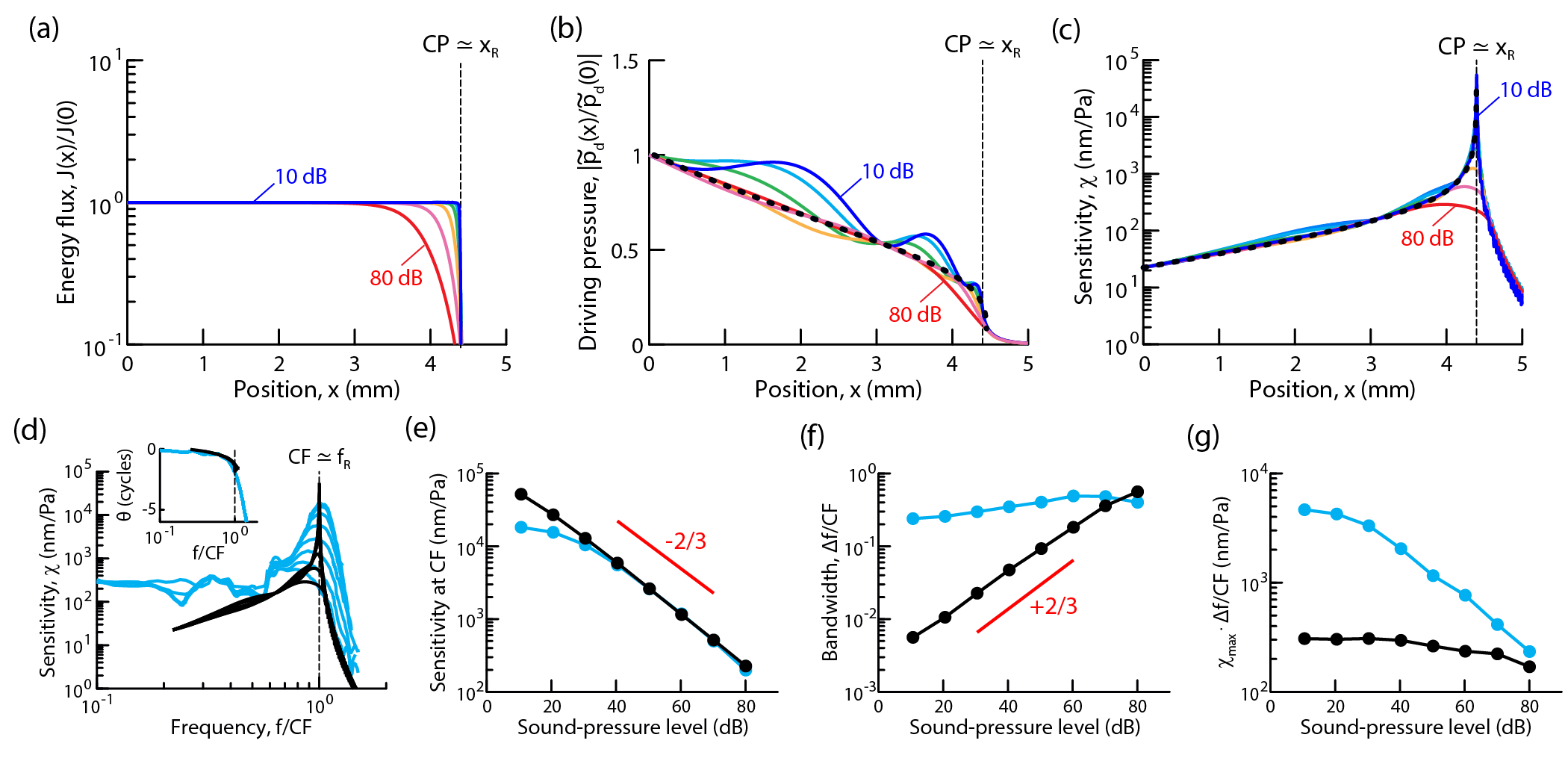} 
\caption{1D model with no energy pumping nor mechanical coupling. (a)--(c) Longitudinal profiles, respectively, of time-averaged energy flux, $J(x)/J(0)$, of driving pressure, $|\tilde{p}_{d}(x)/\tilde{p}_{d}(0)|$, and of sensitivity, $\chi(x)$, for a pure-tone stimulus of frequency $f = 6.6$ kHz at sound-pressure levels increasing from 10 dB (blue) to 80 dB (red) in 10 dB increments. The characteristic place (CP) of peak sensitivity at low sound-pressure levels matches the resonant position, $x_{R}(f)$ of the corresponding oscillator at that position (vertical dashed line). The WKB approximation to the pressure (b) and sensitivity (c) profiles are shown as thick black dashed lines. In (d--g), the results of the model (black) are confronted to experimental measurements (cyan) from Ref.~\cite{Rhode2007}. (d) Sensitivity, $\chi$, and phase ($\theta$, inset) of the response as a function of normalized sound frequency, $f/\textrm{CF}$, for measurements at a fixed position corresponding to the characteristic place (CP) shown in (c) for a tone frequency $f = \textrm{CF} = 6.6$ kHz. The peak sensitivity, $\chi_{\max} = \max(\chi)$, decreases as the sound-pressure level increases from 10 to 80 dB. The characteristic frequency (CF) in the model nearly matches the resonant frequency of the local oscillator: $\textrm{CF} \simeq f_{R}(x = \textrm{CP})$. (e) Sensitivity at CF as a function of sound-pressure level for the data shown in (d); this relationship is well described at all levels by a power law of exponent $-2/3$ (red) in the model and at levels larger than 30 dB in the experiments. (f) Normalized bandwidth, $\Delta f/\textrm{CF}$, of the tuning curves shown in (d) as a function of sound-pressure level; this relationship in the model (black) is well described by a power law of exponent $+2/3$ (red), but not in the experiments. (g) Gain-bandwidth product, $\chi_{\max} \cdot \Delta f/\textrm{CF}$, for the tuning curves shown in (D) as a function of sound-pressure level, in which $\chi_{\max}$ is the peak sensitivity. Parameter values in Table S1~\cite{SuppMat}.}
\label{fig:2}
\end{figure*}

We start with the simplest implementation of a cochlear model based on tonotopically distributed critical oscillators~\cite{Duke2003}. This implementation does not include longitudinal mechanical coupling ($\kappa = 0$ in Eq.~\ref{eqMain:HopfNormalForm}) and is solved using one-dimensional (1D) hydrodynamics: all fields depend only on position $x$ along the longitudinal axis of the box. In addition, with the specific choice $\phi = \pi/2$ of the phase parameter (Eq.~\ref{eqMain:HopfNormalForm}), the oscillators do not provide net energy pumping into the traveling wave (Eq.~\ref{eqMain:ComplexModulusImag}). Accordingly, the fluid-pressure field, $p(x,t)$, and the fluid-velocity field, $u_{x}(x,t)$, evoked by a pure tone (see Section~C2 of the Supplemental Material~\cite{SuppMat}) were associated with a time-averaged energy flux,
\begin{equation}
J = H \langle p(x,t) u_{x}(x,t) \rangle_{t},
\end{equation}
that was uniform along the longitudinal axis of the cochlea at positions where dissipation was negligible (Fig.~\ref{fig:2}(a)). Here the only source of dissipation arises from the cubic nonlinear term in the normal form (Eq.~\ref{eqMain:HopfNormalForm}), which becomes significant when the oscillators' displacement is large enough. For low sound-pressure levels, the energy flux was approximately uniform until it plummeted just before reaching the resonant place, $x_{R}(\omega) = d\log(\omega_{0}/\omega)$, where the natural frequency $\omega_{R}(x_{R}) = \omega$ of the local oscillator matched that of the stimulus (Fig.~\ref{fig:2}(a)); dissipation was thus localized there.

Meanwhile, the pressure driving the oscillators, $\tilde{p}_{d}(x) = \tilde{p}(x)$, decreased progressively toward zero from the base to the resonant place (Fig.~\ref{fig:2}(b)). Under the WKB approximation (derivation in Section~E2 of the Supplemental Material~\cite{SuppMat}), the pressure profile
\begin{equation}
|\tilde{p}(x, \omega)| \simeq |\tilde{p}(0)| \left(\frac{\omega_{R}^{2}(x) - \omega^{2}}{\omega_{R}^{2}(0) - \omega^{2}}\right)^{1/4}
\end{equation}
and the sensitivity
\begin{equation} \label{eqMain:WKBsensitivity_NoPumping}
\chi(x, \omega) \simeq \chi(0, \omega) \left(\frac{\omega_{R}^{2}(0) - \omega^{2}}{\omega_{R}^{2}(x) - \omega^{2}}\right)^{3/4}
\end{equation}
are the same as those obtained when the cochlear partition is described by a string of frictionless harmonic oscillators, which also recalls classic linear models of the cochlea~\cite{Lighthill1981,Zweig1976,Reichenbach2014}. This is because with no energy pumping ($\phi = \pi/2$), the complex modulus $\mathcal{A}(x, \omega) = \bar{\alpha}(\omega_{R}^{2}(x) - \omega^{2})$ of a critical oscillator can be mapped to that of a harmonic oscillator of natural frequency $\omega_{R}$ and mass per unit area $\bar{\alpha}$. We observed pressure profiles that undulated about the WKB approximation (Fig.~\ref{fig:2}(b)); these undulations likely betrayed wave reflections. As the stimulus strength increased, the nonlinear behavior of the oscillators manifested itself: the spatial decay of the energy flux became progressively more graded but the corresponding change in the normalized driving-pressure profiles was moderate.

The sensitivity profiles were well described by their WKB approximation (Eq.~\ref{eqMain:WKBsensitivity_NoPumping}) up to the peak (thick black dashed line in Fig.~\ref{fig:2}(c)) and, correspondingly, sensitivity increased steeply as the wave approached the resonant place $x_{R}(\omega)$. Frequency-tuning curves (Fig.~\ref{fig:2}(d), black) in turn displayed properties that accord with the generic behavior of the critical oscillator at the characteristic place $\textrm{CP}$. The characteristic frequency matched the natural frequency of the local oscillator within less than 0.1\%, corresponding to $f_{R}(\textrm{CP}) \simeq \textrm{CF}$. In addition, the sensitivity at $\textrm{CF}$ evinced a power-law behavior of exponent $-2/3$ (Fig.~\ref{fig:2}(e), black), the bandwidth was inversely related to the sensitivity (Fig.~\ref{fig:2}(f) black) and, correspondingly, the gain-bandwidth product was nearly independent on stimulus level (Fig.~\ref{fig:2}(g), black). 

As a consequence, the model could describe the compressive nonlinearity of experimental level functions relating sensitivity at $\textrm{CF}$ to the sound-pressure level at moderate-to-high levels ($P_{0} > 30$ dB; Fig.~\ref{fig:2}(e)). The resulting tuning curves, however, were too sharp, with a bandwidth only 2\% of that observed in experiments at low sound-pressure levels. In addition, the sensitivity drop beyond CF was much steeper in the model than in experiments. Finally, the model did not exhibit the linear regime of responsiveness observed in experiments at low levels ($P_{0} < 30$ dB; Fig.~\ref{fig:2}(e)), because nothing in the model limits the diverging sensitivity of critical oscillators at vanishing sound-pressure levels.

\subsection{Taking 2D hydrodynamics into account boosts driving pressure and broadens tuning curves but does not decouple sensitivity and tuning}

Cochlear traveling waves are analogous to surface waves~\cite{Lighthill1981}: fluid motion extends transversally ($y$-axis) over a typical distance from the oscillators which is set by the inverse of the wavenumber, $q$. When $qH \ll 1$---the long-wavelength limit, pressure in the fluid, $p(x,y) \simeq p(x)$, is nearly uniform in the transverse direction and hydrodynamics is one-dimensional. Conversely, in the short-wavelength limit, for which $qH \gg 1$, the pressure field, $p(x,y) \simeq p_{d}(x) e^{-qy}$, depends on both coordinates. In the previous section, 1D-hydrodynamics provided a simplifying approximation to solve our cochlear model. The validity of this approximation ($qH \ll 1$), however, breaks down as the wave approaches the resonant place (see Fig.~S1 of the Supplemental Material~\cite{SuppMat}). This is because the wavenumber increases steeply during wave propagation, which results in the progressive confinement of the fluid flow near the oscillators~\cite{Lighthill1981}.

\begin{figure*}[t]
\includegraphics[]{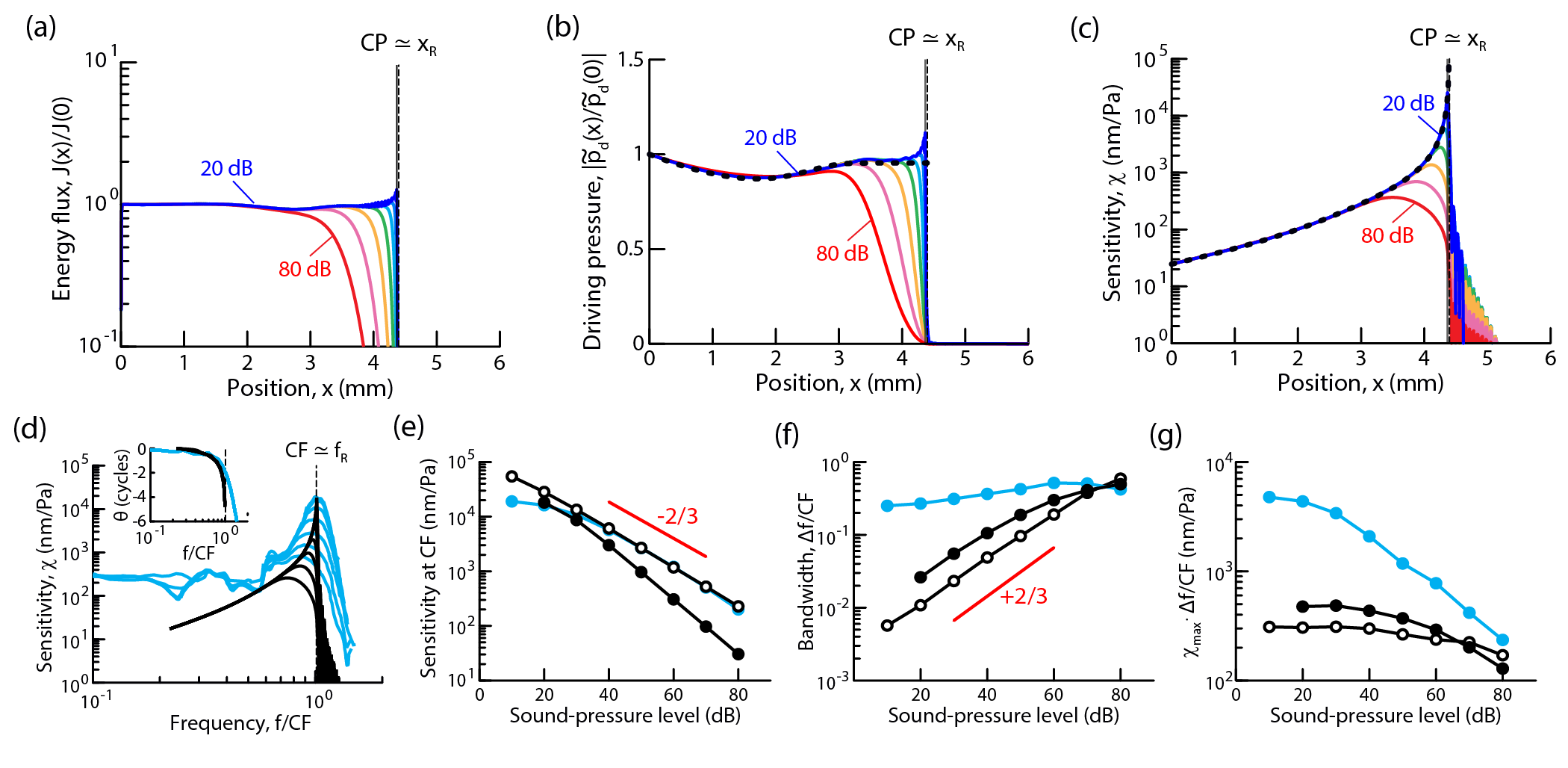}
\caption{2D model with no mechanical coupling nor energy pumping. (a)--(c) Longitudinal profiles, respectively, of time-averaged energy flux, $J(x)/J(0)$, of driving pressure, $|\tilde{p}_{d}(x)/\tilde{p}_{d}(0)|$, and of sensitivity, $\chi(x)$, for a pure-tone stimulus of frequency $f = 6.6$ kHz at sound-pressure levels increasing from 20 dB (blue) to 80 dB (red) in 10 dB increments. The characteristic place (CP) of peak sensitivity at low sound-pressure levels nearly matches the resonant position, $x_{R}(f)$ of the corresponding oscillator at that position (vertical dashed line). The WKB approximation to the pressure (b) and sensitivity (c) profiles are shown as thick black dashed lines. In (d)--(g), the results of the model (black) are confronted experimental measurements (cyan) from Ref.~\cite{Rhode2007} and, in (e)--(g), also to the 1D model (open disks). (d) Sensitivity, $\chi$, and phase ($\theta$, inset) of the response as a function of normalized sound frequency, $f/\textrm{CF}$, for measurements at a fixed position corresponding to the characteristic place (CP) shown in (c) for a tone frequency $f = \textrm{CF} = 6.6$ kHz. The characteristic frequency (CF) in the model still (Fig.~\ref{fig:2}) nearly matches the resonant frequency of the local oscillator: $\textrm{CF} \simeq f_{R}(x = \textrm{CP})$. (e)--(g) Level functions, respectively, of sensitivity at CF, normalized bandwidth, $\Delta f/\textrm{CF}$, and gain-bandwidth product, $\chi_{\max} \cdot \Delta f/\textrm{CF}$, for the tuning curves shown in (d), in which $\chi_{\max}$ is the peak sensitivity. Parameter values in Table S1~\cite{SuppMat}.}
\label{fig:3}
\end{figure*}

At low sound pressure levels, accounting for 2D-hydrodynamics did not change the energy-flux profile (Fig.~\ref{fig:3}(a)) but qualitatively affected how the pressure $p_{d}(x) = p(x,y = 0)$ driving the oscillators (Eq.~\ref{eqMain:HopfNormalForm}) varied with position. In contrast to the gradual decrease toward zero observed with 1D-hydrodynamics (Fig.~\ref{fig:2}(b)), the driving pressure was now approximately uniform before dropping abruptly near the resonant place (Fig.~\ref{fig:3}(b)). In agreement with other cochlear models and called the pressure focusing effect~\cite{Sisto2021}, the driving pressure was thus larger with 2D-hydrodynamics than with 1D-hydrodynamics and this difference increased from the base toward the resonant place. The uniform driving-pressure profile observed with 2D-hydrodynamics can be understood from energy conservation under the WKB approximation, for which $\tilde{u}_{x}(x,y, \omega) \propto q(x, y, \omega) \tilde{p}(x, y, \omega)$ and the time-averaged energy flux
\begin{equation} \label{eqMain:EnergyFlux_2DNoPumping}
J = \int_{0}^{H}\langle p(x,y,t) u_{x}(x,y,t) \rangle_{t} \;dy \propto |\tilde{p}_{d}|^{2}
\end{equation}
is uniform.

As compared to 1D-hydrodynamics (Fig.~\ref{fig:2}), the normalized energy-flux, the normalize driving-pressure and the sensitivity profiles were more sensitive to the sound-pressure level (Figs.~\ref{fig:3}(a)--(c)) with 2D-hydrodynamics. Correspondingly, the tuning curves showed a sensitivity at CF that decreased more steeply as a function of the sound-pressure level (Figs.~\ref{fig:3}(d) and \ref{fig:3}(e), closed disks) and were broader (Figs.~\ref{fig:3}(d) and \ref{fig:3}(f)). Their gain-bandwidth product, however, decreased only weakly with the sound-pressure level (Fig.~\ref{fig:3}(f)). Overall, considering 2D-hydrodynamics did not suffice to strongly decouple sensitivity and bandwidth and thus failed to explain the behavior observed in experiments (Fig.~\ref{fig:3}(d)--(g), cyan).

\subsection{Adding viscoelastic coupling reduces sensitivity and broadens the tuning curves}

With no mechanical coupling between the oscillators, the oscillators' sensitivity profiles plummet beyond the characteristic place and display high curvature at their peak (Figs.~\ref{fig:2}(c) and \ref{fig:3}(c)). This behavior leads to tuning curves that are too sharp as compared with those observed experimentally in the cochlea (Figs.~\ref{fig:2}(d) and \ref{fig:3}(d)).

\begin{figure*}[t]
\includegraphics[]{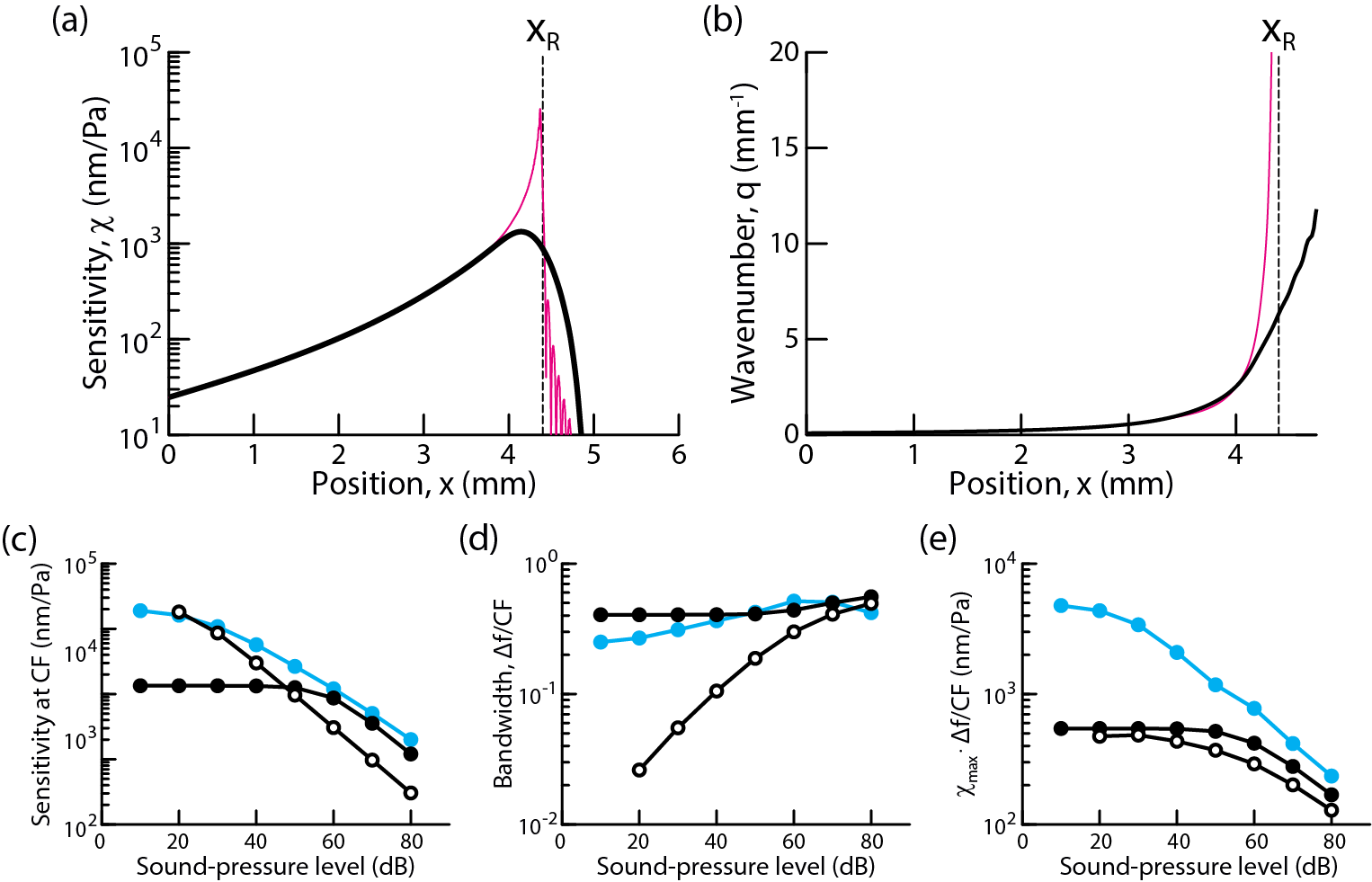}
\caption{2D model with mechanical coupling but no energy pumping. (a) and (b) Longitudinal profiles, respectively, of sensitivity, $\chi(x)$, and wavenumber, $q(x)$, at a sound-pressure level of 20 dB for the 2D model with (thick black line) and without (thin magenta line) coupling. The dashed vertical lines mark the position, $x_{R}(f)$, at which the stimulus frequency matches the natural frequency of the local oscillator. With coupling, the characteristic place of peak sensitivity is basal to the resonant place. (c)--(e) Level functions of sensitivity at $\textrm{CF}$ (c), normalized bandwidth, $\Delta f/\textrm{CF}$, of sensitivity tuning curves (d) and gain-bandwidth product (e) for the 2D model with coupling (closed black disks), without coupling (open black disks, same data as closed black disks in Fig.~\ref{fig:3}(e)--(g)) and in experiments (cyan). Mechanical coupling (black) had both elastic and dissipative components, with $|\kappa_{i}/\kappa_{r}| = 2.5$. Parameter values in Table~S1~\cite{SuppMat}.}
\label{fig:4}
\end{figure*}

Introducing longitudinal viscoelastic coupling ($\kappa \neq 0$ in Eq.~\ref{eqMain:HopfNormalForm}) provided an impediment to curvature of the oscillators' vibration profile, which broadened the peaks and allowed vibrations to propagate beyond the resonant place (Fig.~\ref{fig:4}(a)): oscillators located basally entrained apical ones, generating significant vibration there. Consistent with other cochlear models~\cite{Wickersberg1985,Meaud2010}, coupling elicited a more gradual decline in sensitivity beyond the peak and its dissipative component ($\kappa_{r}$) reduced the peak sensitivity while moving the characteristic place (CP) further toward the base. In addition, toward the resonant place, the wavenumber showed a more gradual rise than the diverging behavior observed with no coupling (Fig.~\ref{fig:4}(b)). 

When the sound-pressure level was varied, we found that coupling gave rise to a linear regime of responsiveness at low-to-intermediate levels ($P_{0} \leq 50$ dB); the sensitivity at CF (Fig.~\ref{fig:4}(c)), the bandwidth (Fig.~\ref{fig:4}(d)) and the gain-bandwidth product (Fig.~\ref{fig:4}(e)) were all independent of sound-pressure level. A linear response occurs at low levels because dissipative coupling effectively generates a non-zero value of the control parameter
\begin{equation} \label{eqMain:EffEpsilon}
\varepsilon = \varepsilon_\mathrm{EFF} \simeq -\kappa_{r}q^{2},
\end{equation}
which effectively detunes the oscillators from criticality. Here the wavenumber, $q = -\partial_{x}\theta$, is set by the longitudinal gradient of the response phase $\theta$. 

Coupling reduced sensitivity to low sound-pressure levels well below experimental values but the bandwidth was now similar in the model and experiments at all levels (Fig.~\ref{fig:4}(c) and \ref{fig:4}(d), cyan). Although mechanical coupling appeared detrimental for auditory detection in terms of sensitivity, it nevertheless afforded a means to broaden and control the shape of frequency-tuning curves, specifically beyond the characteristic frequency.

\subsection{Adding energy pumping into traveling waves boosts sensitivity, revealing nonlinearity while producing level-independent bandwidth}

So far, we have considered critical oscillators with a singular value of the phase parameter $\phi = \pi/2$ in their equation of motion (Eq.~\ref{eqMain:HopfNormalForm}). With this choice, the oscillators neither pump energy into nor dissipate energy from the traveling wave (Eq.~\ref{eqMain:ComplexModulusImag}). We now consider the more general case $0 < \phi < \pi/2$, for which critical oscillators pump energy into the wave as it travels from the base toward the characteristic place. This is the consequence of tonotopy: the traveling wave rides on a string of oscillators of descending natural frequencies and, thus, the driving frequency $\omega$ at each position $x$, is smaller than the natural frequency $\omega_{R}(x)$ of the local oscillator (Fig.~\ref{fig:1}).

\begin{figure*}[t]
\includegraphics[]{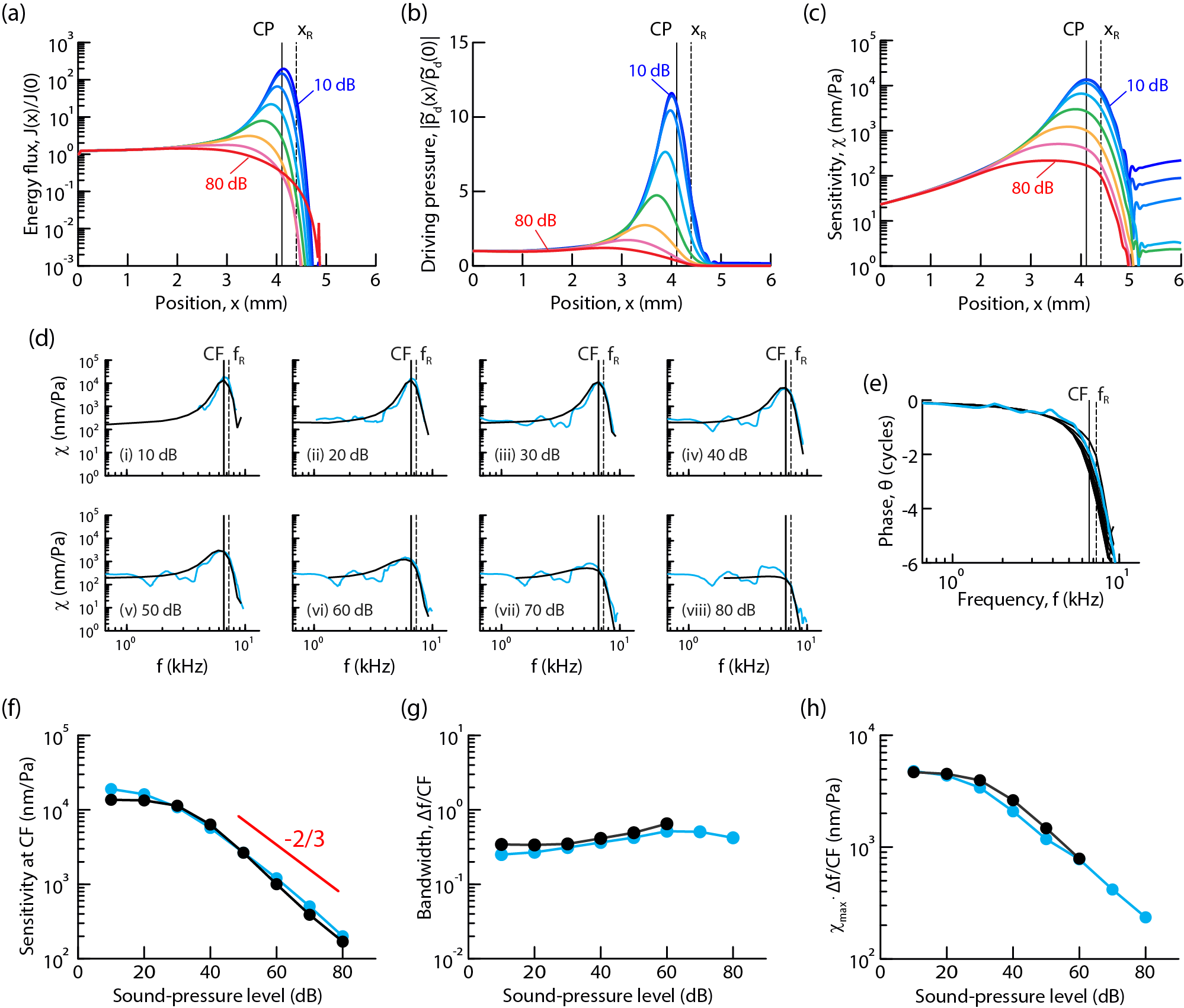}
\caption{2D cochlear model with mechanical coupling and energy pumping---full model. (a)--(c) Longitudinal profiles, respectively, of time-averaged energy flux, $J(x)/J(0)$, of driving pressure, $|\tilde{p}_{d}(x)/\tilde{p}_{d}(0)|$, and of sensitivity, $\chi(x)$, for a pure-tone stimulus of frequency $f = 6.6$ kHz and at sound-pressure levels increasing from 10 dB (blue) to 80 dB (red) in 10 dB increments. The characteristic place, $\textrm{CP} = 4.11$ mm, of maximal sensitivity at low sound-pressure level and the place of oscillator resonance, $x_{R} = 4.40$ mm, with the stimulus are marked by vertical solid and dashed lines, respectively. (d) Comparison of frequency-tuning curves of sensitivity in the model (black) and in experiments (cyan) at a fixed position, $x = \textrm{CP}(f = 6.6$ kHz) and with sound-pressure levels increasing from 10 dB (i) to 80 dB (vi) in 10 dB increments. The characteristic frequency, $\textrm{CF} = 6.6$ kHz, and the resonant frequency, $f_{R} = 7.3$ kHz, of the oscillator at the position of measurement are indicated by vertical solid and dashed lines, respectively. (e) Phase of the response, relative to that at the base ($x = 0$), as a function of frequency for the data shown in (d). (f)--(h) Level functions, respectively, of sensitivity at CF, normalized bandwidth, $\Delta f/\textrm{CF}$, and gain-bandwidth product, $\chi_{\max} \cdot \Delta f/\textrm{CF}$, for the tuning curves shown in (d), in which $\chi_{\max}$ is the peak sensitivity. In (d)--(h), the results of the model (black) are confronted to experimental measurements (cyan) from Ref.~\cite{Rhode2007}. Parameter values in Table~S1~\cite{SuppMat}.}
\label{fig:5}
\end{figure*}

Energy pumping by the individual oscillators resulted in the progressive increase of the energy flux (Fig.~\ref{fig:5}(a)) and of the driving pressure (Fig.~\ref{fig:5}(b)) during propagation of the traveling wave, corresponding to spatial accumulation of energy gain. Near the characteristic place, dissipative coupling between the oscillators and nonlinear dissipation by the oscillators themselves (Eq.~\ref{eqMain:HopfNormalForm}) stopped further increase and led to prominent peaks in the energy-flux and driving-pressure profiles. In striking contrast to models with no energy pumping (Figs.~\ref{fig:2}(b) and \ref{fig:3}(b)), the driving pressure could be much larger at the characteristic place than at the base, here by about tenfold for a sound-pressure level of 30 dB (Fig.~\ref{fig:5}(b)). Amplification of the driving pressure through energy pumping into the traveling wave boosted sensitivity to weak stimuli (compare Fig.~\ref{fig:4}(a), black and Fig.~\ref{fig:5}(c), blue). The characteristic place ($\textrm{CP} < x_{R}$) of peak sensitivity remained well distinct from the position of local mechanical resonance ($x = x_{R}$) (Fig.~\ref{fig:5}(c)), a basal shift that could be attributed to dissipative coupling (Fig.~\ref{fig:4}(a)).

Altogether, the interplay between the generic properties of critical oscillators, mechanical coupling and traveling waves produced level-dependent tuning curves of sensitivity (Fig.~\ref{fig:5}(d), black), as well as the relation between the response phase and frequency (Fig.~\ref{fig:5}(e), black), that could match experimental measurements over a broad range of sound-pressure levels (Fig.~\ref{fig:5}(d) and (e), cyan; details of the matching procedure in Section~D3 of the Supplemental Material~\cite{SuppMat}). Energy pumping counteracted the detrimental effect of mechanical coupling on sensitivity (Fig.~\ref{fig:4}(b)) by extending toward low sound-pressure levels a compressive nonlinearity (Fig.~\ref{fig:5}(f), black) that could here only stem from the nonlinear behavior of the critical oscillators. This nonlinearity was well described by a power law of exponent $-2/3$, but not precisely: a power-law fit led to an exponent of $-0.80$, thus smaller than that for a critical oscillator in isolation. Although the peak sensitivity decreased by about two orders of magnitude for sound-pressure levels that increased from 10 dB to 80 dB, the bandwidth increased only little over the same range, by about twofold (Fig.~\ref{fig:5}(g), black). Accordingly, the gain-bandwidth product decreased strongly with level, by more than tenfold (Fig.~\ref{fig:5}(h), black).

\subsection{ A minimal, 1D effective model captures the active nonlinear behavior of the cochlea}

We have shown that energy accumulation due to energy pumping into the traveling wave leads to prominent peaks in the driving-pressure profiles (Fig.~\ref{fig:5}(b)). This observation suggests that the pressure-focusing effect resulting from 2D-hydrodynamics, which, by itself, cannot produce driving-pressure peaks (Eq.~\ref{eqMain:EnergyFlux_2DNoPumping}; Fig.~\ref{fig:3}(b)), may be dominated by the effects of energy pumping on pressure. To test this inference, we returned to the model under the approximation of 1D-hydrodynamics while keeping energy pumping and mechanical coupling. We found level-dependent tuning curves that still compared favorably with experiments (Fig.~S2)~\cite{SuppMat}. We note that some parameter values had to be readjusted (Table~S1)~\cite{SuppMat}. Specifically, the phase parameter $\phi$ was farther from the value $\pi/2$ to compensate the loss of the pressure focusing effect afforded by 2D hydrodynamics by increasing energy pumping.

Simplifying the model by using the 1D-approximation is advantageous, for it provides a computationally efficient description of cochlear tuning curves and is amenable to analytics. Under the WKB approximation (see derivation in Section E2 of the Supplemental Material~\cite{SuppMat}), the sensitivity
\begin{equation} \label{eqMain:GainFactor}
\chi = \chi_{\varnothing} \cdot \mathcal{G},
\end{equation}
can be written as the product of the sensitivity in the absence of energy pumping ($\phi = \pi/2$), $\chi_{\varnothing}$ (given by Eq.~\ref{eqMain:WKBsensitivity_NoPumping}), and a gain factor, $\mathcal{G}$, resulting from progressive energy accumulation into the traveling wave. Using $\bar{\omega}(x) = \omega/\omega_{R}(x)$, this factor can be expressed as
\begin{equation}
\mathcal{G}^{2}(x) = \left(\sin^{2}(\phi) + \bar{\omega}^{2}(x)\cos^{2}(\phi)\right)\frac{J(x)}{J(0)},
\end{equation}
in which the energy gain
\begin{equation} \label{eqMain:EnergyGain}
\frac{J(x)}{J(0)} = \exp\left[4N_{\varnothing} \frac{\cos(\phi)}{\sqrt{\sin(\phi)}}\left(1 - \sqrt{1 - \bar{\omega}^{2}(x)}\right)\right]
\end{equation}
depends on only two independent dimensionless parameters: the phase parameter $\phi$ and the number of cycles, $N_{\varnothing} = \frac{d}{4}\sqrt{\frac{2\rho_{0}}{\bar{\alpha}H}}$, that the wave accumulates to reach the resonant place when there is no energy pumping; in the presence of pumping, the number of cycles is reduced to $N = N_{\varnothing}\sqrt{\sin(\phi)} < N_{\varnothing}$.

\begin{figure}[t]
\includegraphics[]{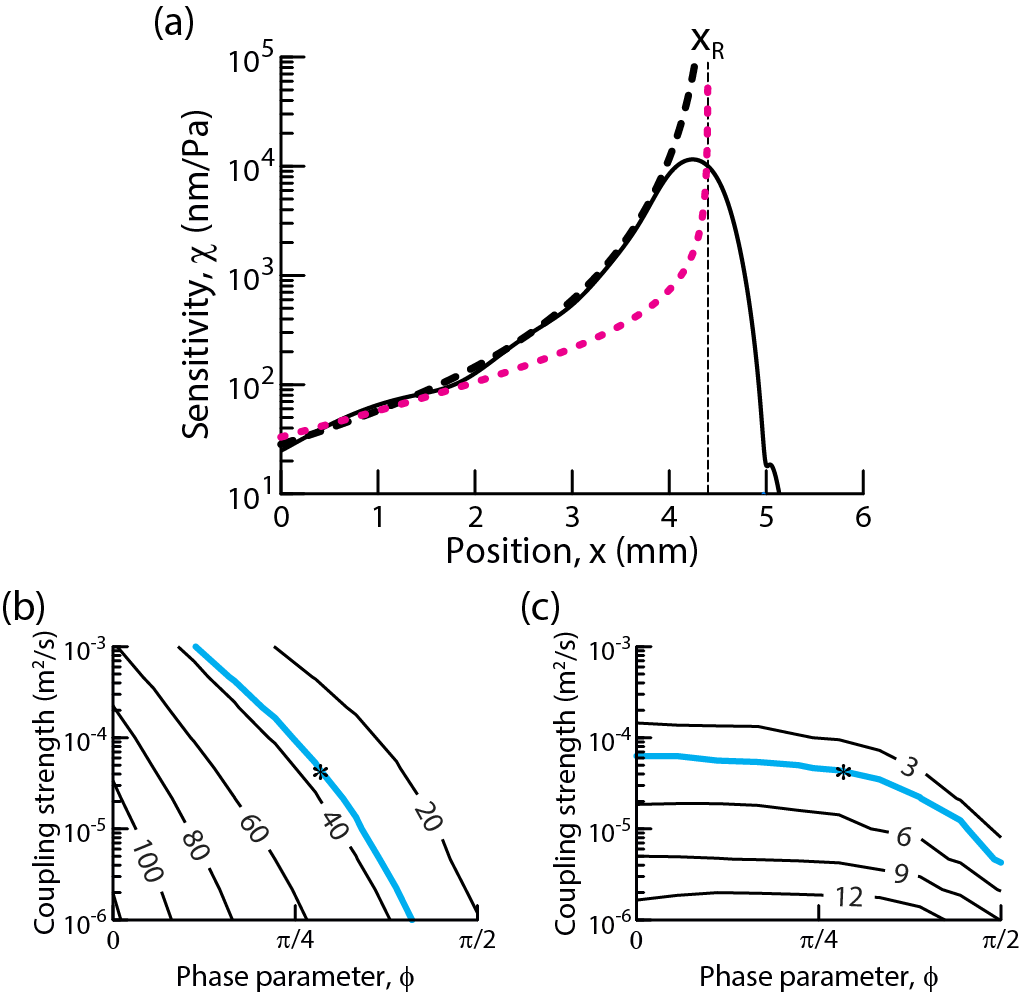}
\caption{Differential control of sensitivity and bandwidth by energy pumping and mechanical coupling. (a) Longitudinal profile of sensitivity in response to a 30-dB stimulus at 6.6kHz with pumping and coupling (black continuous line). Its WKB approximation (thick black dashed line) is compared to that when there is no pumping (magenta); these approximations are given by $\chi(x)$ in Equations~\ref{eqMain:GainFactor}--\ref{eqMain:EnergyGain} and by $\chi_{\varnothing}(x)$ in Equation~\ref{eqMain:WKBsensitivity_NoPumping}, respectively. The vertical dashed line shows the resonant place, $x_{R}$. (b) Lines of equal sensitivity in the plane ($\kappa_{r}(x = 0), \phi$), where $\kappa_{r}(x = 0) = \bar{\kappa}\omega_{0}$ ($\omega_{0} = 1.89 \times 10^{5}$ rad s$^{-1}$ is fixed, $\bar{\kappa}$ varies) and $\phi$ are the parameters that control the strength of dissipative coupling and energy pumping, respectively. Sensitivity, $\chi(\textrm{CF})/\chi(f \rightarrow 0)$, was measured at CF and normalized by sensitivity at low frequencies; the values indicated on the isolines are in decibels. (c) Lines of equal quality factor in the plane ($\kappa_{r}(0), \phi$). The quality factor was defined as $\textrm{CF}/\Delta f$, where $\textrm{CF}$ is the characteristic frequency and $\Delta f$ is the bandwidth of the tuning curve, which was measured 10 dB below the peak. In (b) and (c), the cyan line shows, respectively, the value of normalized sensitivity and quality factor measured in experiments at 20 dB~\cite{Rhode2007}, which are 35 dB and 4, respectively. The stars mark the point ($\phi = \pi/2 - 0.67$, $\bar{\kappa}_{r} = 224$ $\mu$m$^{2}$) where the model matches both experimental values. Parameter values in Table S1~\cite{SuppMat}.}
\label{fig:6}
\end{figure}

Approaching the resonant place ($\bar{\omega}(x) \rightarrow 1$) leads to a power-law divergence of $\chi_{\varnothing}$ (Eq.~\ref{eqMain:WKBsensitivity_NoPumping}). Far enough from the base, energy pumping into the traveling wave boosts sensitivity ($\mathcal{G} > 1$), which can reach relatively high levels even away from the resonant place. At the characteristic place $\textrm{CP} = 4.1$ mm, sensitivity peaks and $\mathcal{G}(\textrm{CP}) = 64$ (Fig.~\ref{fig:6}(a)). We observed that the WKB approximation captures the sensitivity of tuning curves for frequencies up to the peak (thick black dashed line in Fig.~\ref{fig:6}(a)). The growth in sensitivity stops following its WKB approximation when dissipation becomes significant. At the peak, as discussed above in the case of 2D hydrodynamics (Fig.~\ref{fig:4}(a)), mechanical coupling limits curvature of the sensitivity profile; beyond the peak, it also controls the graded drop in sensitivity.

Importantly, in the linear regime of responsiveness, energy pumping and mechanical coupling were both essential to describe the peak sensitivity relative to that at low frequencies, $\chi(\textrm{CF})/\chi(f \rightarrow 0)$, and the quality factor, $\textrm{CF}/\Delta f$, of experimental tuning curves, in which $\Delta f$ is the bandwidth. Indeed, by systematically exploring how energy pumping (parameter $\phi$) and dissipative coupling (parameter $\kappa_{r}$) affected these two key features of the tuning-curve shape (Figs.~\ref{fig:6}(b) and \ref{fig:6}(c)), we found that isolines of sensitivity were steeper than those of quality factor: increasing energy pumping by decreasing $\phi$ resulted in a larger relative change in sensitivity than in quality factor. Consequently, there was a single pair of parameters, (e.g. $\phi = \pi/2 - 0.67$ and $\bar{\kappa}_{r} = 224$ $\mu$m$^{2}$ in Fig.~\ref{fig:6}(b) and (c)) for which the model could match the experimental values of sensitivity and quality factor.

\begin{figure}[t]
\includegraphics[]{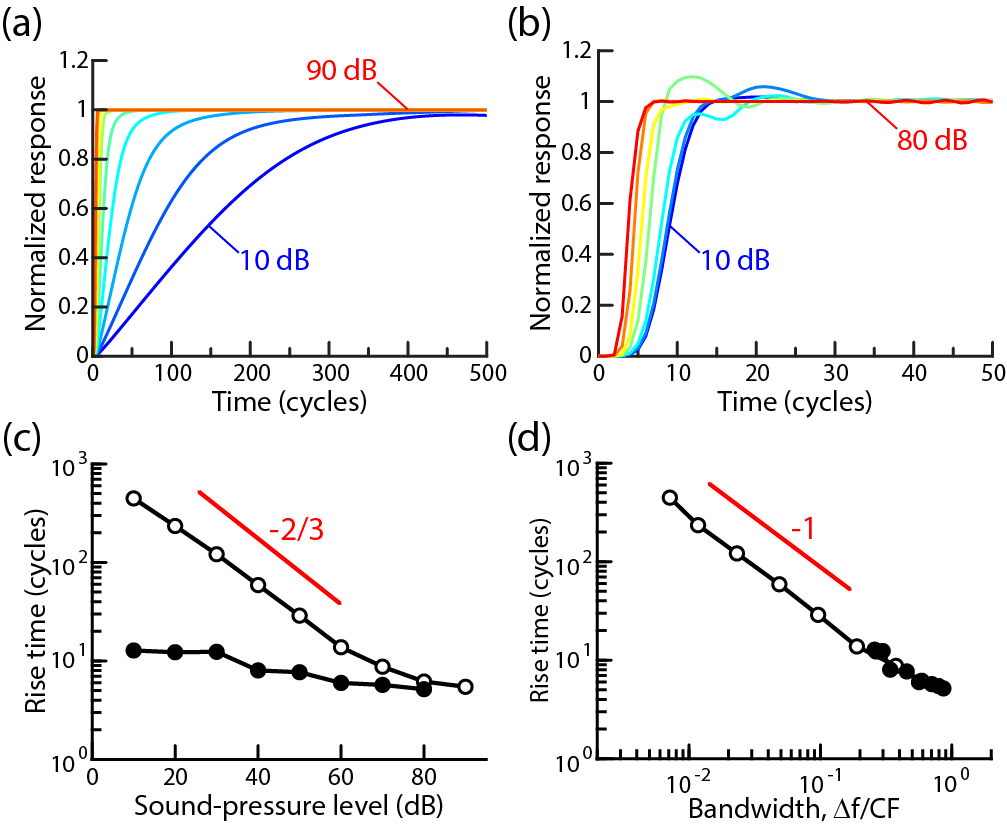}
\caption{Effects of mechanical coupling and energy pumping on response time. (a) Time course of the response to a 6.6-kHz stimulus starting at the time origin for sound-pressure levels increasing from 10 dB (blue) to 90 dB (red) in 10-dB increments with no mechanical coupling between the oscillators nor energy pumping by the oscillators into the traveling waves (same data as for Fig.~\ref{fig:2}). (b) Same as in (a) but with mechanical coupling and energy pumping (same data as for Fig.~S2)~\cite{SuppMat}. (c) Rise time as a function of sound-pressure level for the responses shown in (a) (open disks) and (b) (closed disks). (d) Rise time as a function normalized bandwidth of tuning curves for the responses shown in (a) (open disks) and (b) (closed disks). For each stimulus, the rise time was measured at 95\% of the steady-state response. Parameter values in Table~S1~\cite{SuppMat}.}
\label{fig:7}
\end{figure}

Finally, our time-domain resolution of the cochlear model allowed us to compute the rise time to a steady-state response upon application of a pure tone stimulus, here at a frequency of 6.6 kHz. With no mechanical coupling nor energy pumping, the high sensitivity and sharp frequency tuning to low levels (Fig.~\ref{fig:2}(d)--(f)) were associated with long rise times of several hundred of cycles (Fig.~\ref{fig:7}(a)). Matching experimental tuning curves by adding coupling and pumping (Fig.~S2)~\cite{SuppMat} led to rise times that were much smaller, only about ten cycles (Fig.~\ref{fig:7}(b)). While the rise time decreased steeply with level in the first case---as the power law of exponent $-2/3$ expected from critical slowing down acting locally (see Section~B of the Supplemental Material~\cite{SuppMat}), it remained nearly independent of level when coupling and pumping were added (Fig.~\ref{fig:7}(c)). These behaviors reflect an inverse relation between the bandwidth of frequency tuning and the rise time in our nonlinear model (Fig.~\ref{fig:7}(d)).

\section{Discussion}

Operation at a Hopf bifurcation provides an appealing principle for auditory detection at a characteristic frequency of sound. As in other manifestations of critical phenomena, this principle ensures uttermost sensitivity to weak external stimuli and, in turn, an extended dynamical range of responsiveness. A critical oscillator is also a resonant system, which responds most sensitively near a characteristic frequency and, thus, provides frequency selectivity. However, criticality also results in ``critical slowing down'', by which the increasing sensitivity to smaller stimulus levels comes with an increasing rise time to reach a steady-state response. As other resonators, a critical oscillator shows an inverse relation between sensitivity and the bandwidth of its frequency tuning or, equivalently, a proportional relation between sensitivity and response time. Its generic nonlinear behavior dictates that peaks of sensitivity in frequency-tuning curves are ``small and broad'' at high levels and progressively ``taller and thinner'' at decreasing levels, leading to long response times for weak stimuli. In contrast to a critical oscillator, the cochlea decouples how sensitivity and bandwidth of frequency tuning vary with the sound-pressure level. This way, the cochlea manages to operate over a wide dynamical range of sound-pressure levels while ensuring that its response time remains relatively short and level independent.

Our work demonstrates that cochlear models based on a set of tonopically-distributed critical oscillators can overcome the inverse relation between sensitivity and bandwidth imposed on their individual constituents. We found that sensitivity to increasing sound-pressure levels displayed a compressive nonlinearity that preserved, but not precisely (see below), the generic power-law behavior of a critical oscillator while the bandwidth of frequency tuning remained broad and increased only little with level. As a result, the model could account, with a single set of parameters, for the shape of cochlear tuning curves over a broad range of sound-pressure levels (Fig.~\ref{fig:5}(d)). This finding alleviates a major criticism addressed to criticality as the fundamental basis of nonlinear cochlear amplification (see e.g.~\cite{Lyon2017,Kondylidis2025}).

\subsection{Local criticality vs. distributed effects from mechanical coupling and energy pumping}

With no mechanical coupling between the oscillators nor energy pumping by the oscillators into the traveling waves, peak and bandwidth of nonlinear tuning curves are dictated by local resonance of a critical oscillator (Fig.~\ref{fig:2}) and, thus, the modeled cochlea suffers from critical slowing down (Fig.~\ref{fig:7}(a)). With mechanical coupling, the peak response is instead shaped by an ensemble of oscillators driven off resonance, which broadens the tuning curves, but coupling also introduces a linear regime of reduced sensitivity so that the relation between sensitivity and bandwidth remains flat (Fig.~\ref{fig:4}). Energy pumping can compensate for the loss in sensitivity imposed by coupling while keeping a broad and level-independent bandwidth (Figs.~\ref{fig:5}, \ref{fig:6} and Fig.~S2)~\cite{SuppMat}. Pumping results in spatial accumulation of energy gain as the wave travels toward the characteristic place while, without it, the increase in sensitivity relies solely on local resonance (Fig.~\ref{fig:6}(a)).

Decoupling sensitivity and bandwidth results from non-local interactions between the oscillators via mechanical coupling and energy pumping into traveling waves. The cochlear model works as a distributed system that benefits from an interplay between local criticality and waves to operate over a large dynamical range of sound-pressure levels (Figs.~\ref{fig:5} and S2) without suffering from critical slowing down (Fig.~\ref{fig:7}).

\subsection{On the origin of the cochlear nonlinearity}

Our cochlear model distinguishes itself from other models by employing the Hopf normal form (Eq.~\ref{eqMain:HopfNormalForm}) to describe the local transverse motion of the basilar membrane. The normal form is the mathematical consequence of the principle of local operation of each oscillator at a Hopf bifurcation. Importantly, the dominant nonlinearity in the normal form is cubic and phase invariant; there is no quadratic term (Eq.~\ref{eqMain:HopfNormalForm}). This nonlinearity is generic; it emerges robustly irrespective of the detailed mechanism that locally brings the system at the bifurcation. There is thus no choice for the nonlinearity of the local oscillator in the traveling-wave box-model of the cochlea developed here.

The compressive nonlinearity that emerged from collective interactions between the critical oscillators was not strictly dictated by the power-law behavior associated to criticality of individual oscillators: this is because the cochlea, as a whole, is not critical! A power-law fit to the relation between sensitivity at the characteristic frequency and sound-pressure level in the ear canal indeed led to an exponent $-0.80$, thus close to, but smaller than, the generic value of $-2/3$ expected from a single oscillator driven in isolation (Fig.~\ref{fig:5}). In addition, the overall power-law exponent of the distributed system of coupled oscillators also varied depending on parameter values, for instance of parameter $\beta_{i}$ (Fig.~S3A)~\cite{SuppMat}: it is not universal. In experiments~\cite{Dewey2023}, the exponents to power-law fits to level functions are also variable and smaller than $-2/3$ (Fig.~S3B)~\cite{SuppMat}). Although this property might be mistaken for evidence against an underlying principle of criticality, our results show that it is in fact consistent with this hypothesis. By accounting for the complex nonlinear behavior of cochlear mechanics, the interplay between the generic nonlinear behavior of critical oscillators---the only nonlinearity in the model---and traveling waves captured the essence of the cochlear nonlinearity.

In detailed descriptions of the cochlea, nonlinearity typically stems from the sigmoidal relation between the transduction current and deflection of the hair bundle---the mechanical antenna of the sensory hair cells~\cite{Ramamoorthy2007,Samaras2023,Bowling2021}. Note, however, the passive and broadband nonlinearity of the transducer cannot, by itself, account for the cochlear nonlinearity~\cite{Barral2012}, which is an emergent property of the system as a whole. Detailed descriptions of the dynamic interplay between mechano-electrical transduction by the hair cells, micromechanics of the organ of Corti, force feedback by outer-hair-cell electromotility, and hair-cell coupling via cochlear fluids and viscoelastic membranes produce nonlinear tuning curves that can match experimental measurements~\cite{Meaud2010,Ramamoorthy2007,Samaras2023}. In other, simpler models that can also describe experiments, nonlinearity is introduced phenomenologically by assuming that a single parameter, usually associated with the active process, varies with sound-pressure level---turning a knob~\cite{Sisto2021,Altoe2020}.

\subsection{Critical or not?}

We propose that successful cochlear models match the observed nonlinearity because they effectively tune each cross-section of the model in the proximity of a Hopf bifurcation, or they mimic the generic nonlinearity of critical oscillators. Three lines of evidence support this conjecture.

First, some cochlear models have been shown to become unstable via a Hopf bifurcation~\cite{Samaras2023,Nankali2015}.

Second, the principle of criticality ensures that the critical oscillators pump energy into the traveling wave when they are driven below their characteristic frequency~\cite{Hudspeth2010,Martin2021} (Eq.~\ref{eqMain:ComplexModulusImag}). Thus, a traveling wave riding on a tonotopic organization of critical oscillators with descending natural frequencies automatically benefits from accumulation of energy gain. This property accounts for direct and indirect experimental evidence indicate that the organ of Corti pumps energy into the traveling wave within an extended region basal to the characteristic place~\cite{Hudspeth2014,Zweig1991,Shera2007,Dong2013,Fisher2012}.

Third, inference of the local equation of basilar-membrane motion from experimental tuning curves has revealed that the basilar membrane behaves as a string of uncoupled harmonic oscillators driven by a linear combination of pressure and its time derivative~\cite{Altoe2020,Zweig2015}. The inferred equation of motion, remarkably enough, accords with the generic linear response of a critical oscillator (Eq.~\ref{eqMain:HopfNormalForm} with $\beta = 0$ and $\kappa = 0$)
\begin{equation} \label{eqMain:LinearHopfVsZweig}
\begin{aligned}
\ddot{h}-2\varepsilon\dot{h}+(\omega_R^2+\varepsilon^2)h
&= \frac{\omega_R}{\alpha}\left[
    \left(\sin\phi-\frac{\varepsilon}{\omega_R}\cos\phi\right)p_d
\right.\\
&\qquad\left.
    +\frac{\cos\phi}{\omega_R}\dot{p}_d
\right],
\end{aligned}
\end{equation}
with $\varepsilon/\omega_{R} = \bar{\varepsilon} \simeq -0.03$ and $\phi \simeq 0.75$ rad~\cite{Zweig2015} (see mapping in Section~F of the Supplemental Material~\cite{SuppMat}). Thus, the oscillator satisfies the necessary condition for energy pumping ($\phi \neq \pi/2$, Eq.~\ref{eqMain:ComplexModulusImag}) and, though not precisely critical, is only slightly detuned from the critical point ($|\varepsilon| \ll \omega_{R}$ and $\varepsilon < 0$).

Criticality betrays itself when the value of parameter $\varepsilon$ is small enough to reveal the cubic nonlinearity over an extended range of physiological stimulus levels. Is the value of $\varepsilon$ given above small enough to qualify the oscillator as a critical oscillator? To address this question, we used our nonlinear cochlear model in the case where the oscillators are detuned from criticality ($\varepsilon = \bar{\varepsilon}\omega_{R}(x)$ with $\bar{\varepsilon} \neq 0$ in Eq.~\ref{eqMain:HopfNormalForm}) and mechanically uncoupled ($\kappa = 0$), as suggested by Equation~\ref{eqMain:LinearHopfVsZweig}. Confronting this realization of the model to experiments is expected to provide an estimate of the distance to criticality.

We found that a string of uncoupled, but non-critical, oscillators could also lead to a compressive nonlinearity in accord with that in experiments over at least two orders of magnitude of sound-pressure levels with $\varepsilon/\omega_{R} \simeq -0.09$ and $\phi = 0.57$ rad (Fig.~S4)~\cite{SuppMat}: a value $|\varepsilon/\omega_{R}| < 0.1$ is small enough to preserve the nonlinear behavior associated with criticality.

Thus, the oscillators do not need to be precisely critical to elicit a compressive nonlinearity over a broad range of stimulus levels. As already noted above (Eq.~\ref{eqMain:EffEpsilon}), critical oscillators that are mechanically coupled ($\varepsilon = 0$ and $\kappa \neq 0$) effectively behave as uncoupled oscillators that are detuned from criticality ($\varepsilon \neq 0$ and $\kappa = 0$) with an effective control parameter $\varepsilon_\mathrm{EFF} \simeq -\kappa_{r}q^{2}$, where $\kappa_{r} = \bar{\kappa}_{r}\omega_{R}$ is the strength of dissipative coupling and $q$ is the wavenumber. In support of this interpretation, with $\bar{\kappa}_{r} = 160$ $\mu$m$^{2}$ and $q(\textrm{CP}) = 17.8$ rad mm$^{-1}$ at the characteristic place (Fig.~5), we find $\bar{\varepsilon}_\mathrm{EFF} = \varepsilon_\mathrm{EFF}/\omega_{R}(\textrm{CP}) = -0.05$, which is consistent with the other estimates of the control parameter given above. The intrinsic value of $|\bar{\varepsilon}|$ for individual oscillators must be smaller than $|\bar{\varepsilon}_\mathrm{EFF}| \simeq 0.1$, for mechanical coupling can only increase the effective value of the control parameter. Although we did not take them into account, we note that fluctuations also effectively detune critical oscillators~\cite{Julicher2009,Nadrowski2004} and, thus, the requirement of a small $|\varepsilon|$ is expected to be even more stringent to maintain the effective value at $|\bar{\varepsilon}_\mathrm{EFF}| \simeq 0.1$.

In our cochlear model, by effectively detuning the oscillators from criticality, coupling limits the normalized peak sensitivity of a single oscillator to an effective value, $\frac{\chi(\omega_{R})}{\chi(\omega = 0)} \simeq \frac{1}{|\bar{\varepsilon}_\mathrm{EFF}|\sin(\phi)} = 22$, using $\bar{\varepsilon}_\mathrm{EFF} = -0.05$ and $\phi = \pi/2 - 0.4$ (see above and parameter values for Fig.~\ref{fig:5} in Table~S1)~\cite{SuppMat}. This value is about a third of the sensitivity achieved at the characteristic place from energy accumulation by tonopically-distributed oscillators (Fig.~\ref{fig:5}). Spatial accumulation of energy gain during propagation of the traveling wave mitigates the need to locally operate precisely at a critical point to achieve high sensitivity to low sound-pressure level~\cite{Reichenbach2010}.

In this wave-centric view, as recognized earlier~\cite{Lyon1988,Kern2003}, cascading the modest gains of uncoupled oscillators operating away from a critical point may account for high gain while allowing for the relatively broad tuning curves observed in the cochlea at low sound-pressure levels. We emphasize that longitudinal coupling between the oscillators requires that the oscillators operate closer to criticality---oscillators are precisely critical in our description of the cochlea---to produce similar tuning curves as with no coupling (Fig.~\ref{fig:6} and Fig.~S4)~\cite{SuppMat}; neglecting coupling thus leads to an overestimation of the distance to the critical point. In addition, the wave-centric view of cochlear amplification does not specify why the cochlear nonlinearity extends over three to four orders of magnitude of sound-pressure levels nor how proper energy pumping comes about. In contrast, the principle of criticality specifies both the physical nature of the nonlinearity and that of energy pumping.

\section{Conclusion}

Our work has revealed how the interplay between local nonlinear properties of critical oscillators and nonlocal effects from traveling waves shapes cochlear tuning curves. It reconciles seemingly opposite views on cochlear amplification that are based on local amplification by critical oscillators on the one hand and distributed energy accumulation due to wave propagation on the other hand. Maintaining relatively broad and level-independent tuning curves while amplifying weak sound stimuli is essential to endow the cochlea with a broad dynamical range of responsiveness and a relatively fast responsiveness to weak stimuli at all levels. Marrying tonopically distributed critical oscillators with traveling waves provides a fundamental physical principle to explain nonlinear sound processing in the cochlea.

\begin{acknowledgments}
This research was supported by the Fondation pour l'Audition (project n°FPA RD 2020-7) and the Labex Cell(n)Scale (ANR-11-LABX-0038 and ANR-10-IDEX-001-02). Henri Ver Hulst is an alumni of the Paris Sciences et Lettres University and acknowledges a grant from the doctoral school ``Ecole Doctorale Paris Ile de France'' (EDPIF). PM thanks Richard F. Lyon for pointing out that critical oscillators lead to way-to-sharp responses with respect to the cochlea and Jim Hudspeth for inspiring discussions on criticality and hearing over the past 28 years.
\end{acknowledgments}


\end{document}